\documentclass[reprint,superscriptaddress
]{revtex4-2}

\usepackage{amsmath}
\usepackage{amsfonts}
\usepackage{hyperref}
\usepackage{amssymb}
\usepackage{graphicx}
\usepackage{hyperref}
\usepackage{colortbl}
\usepackage{subfigure}
\usepackage{gensymb}
\usepackage{svg}

\usepackage[T1]{fontenc}
\usepackage{mathptmx}
\usepackage{bbold}
\usepackage{scalerel}
\usepackage{xcolor}
\usepackage{mathtools}
\usepackage{epstopdf}

\newcommand*\ch[1]{\ensuremath{\mathrm{#1}}}
\newcommand{\hyref}[1]{\hyperref[#1]{\ref{#1}}}

\newcommand{\orange}[1]

\newcommand{\zbar}{\raisebox{0.2ex}{--}\kern-0.6em Z}

\begin{document}

 	\title{Coulombic surface-ion interactions induce non-linear and chemistry-specific charging kinetics}
\author{W.Q. Boon}
	\affiliation{Institute for Theoretical Physics, Utrecht University,  Princetonplein 5, 3584 CC Utrecht, The Netherlands}
	\author{M. Dijkstra}
	\affiliation{Soft Condensed Matter, Debye Institute for Nanomaterials Science, Utrecht University, Princetonplein 1, 3584 CC Utrecht, The Netherlands}
	\author{R. van Roij}
		\affiliation{Institute for Theoretical Physics, Utrecht University,  Princetonplein 5, 3584 CC Utrecht, The Netherlands}
	
 \begin{abstract}
	While important for many industrial applications, chemical reactions responsible for charging of solids in water are often poorly understood. We theoretically investigate the charging kinetics of solid-liquid interfaces, and find that the time-dependent equilibration of surface charge contains key information not only on the reaction mechanism, but also on the valency of the reacting ions. We construct a non-linear differential equation describing surface charging by combining chemical Langmuir kinetics and electrostatic Poisson-Boltzmann theory. Our results reveal a clear distinction between late-time (near-equilibrium) and short-time (far-from-equilibrium) relaxation rates, the ratio of which contains information on the charge valency and ad- or desorption mechanism of the charging process. Similarly, we find that single-ion reactions can be distinguished from two-ion reactions as the latter show an inflection point during equilibration. Interestingly, such inflection points are characteristic of autocatalytic reactions, and we conclude that the Coulombic ion-surface interaction is an autocatalytic feedback mechanism.
	\end{abstract}

 \maketitle

		Charged solid-liquid interfaces play a central role in a wide variety of industries such as food and coating production \cite{food3,paint1,paint2}, mining \cite{mine1,mine2,mine3}, medicine \cite{medicine1,medicine2,medicine3}, soil remediation \cite{soil1,soil2, pollutant} and even carbon capture \cite{carboncapture}. With the advent of nanoscale fluidics one expects that charged surfaces become ever more important
		\cite{nano1, nano2}. In water and other polar solvents chemical reactions are a common mechanism by which surfaces obtain their charge. For ionic solids the de- or adsorption of a dissolved ionic compound is often preferred over the sorption of its own counterion  \cite{database, review_electrification, review_electrification_2}; for covalent solids such as polymers and metal  oxides the acidic nature of surface groups ensures that the surface (de)protonates in polar solvents and hence becomes charged \cite{database_unknown_theory,polymer,polymer2, review_electrification, review_electrification_2}. However, for many processes of industrial and environmental importance relatively  little is known about the surface chemistry \cite{database_unknown_theory, review_electrification, review_electrification_2} as the electrolytes in realistic applications contain a large variety of ions that can all undergo multiple reactions \cite{review_electrification, review_electrification_2, 2pK_2}. Due to experimental limitations the majority of studies investigating surface charging are performed at (quasi)-equilibrium conditions \cite{review_electrification, review_electrification_2}, with the notable exception of pressure-jump experiments \cite{pressure_jump, pressure_jump2}. Only recently, however, it has been shown that the kinetics of chemical surface reactions can strongly couple to electrokinetic fluid flows, thereby affecting the physical surface properties on macroscopic scales  \cite{chem_macr2,chem_macr3,chem_macr4,hartkamp,SFG8, review_electrification_2}. Furthermore, with the recent advent of fast and surface specific non-linear spectroscopy the dynamic measurement of surface charge has become feasible \cite{SFG3,SFG4,SFG5,SFG6,SFG7, SFG8}. In this context it has been explicitly stated that there is an urgent need for theoretical models to describe such experiments \cite{database_need_theory}. Traditionally, sorption kinetics is typically described by (pseudo)-first-order reactions \cite{pollutant, first_order, relaxation_methods} that exhibit single-exponential relaxation towards equilibrium; the influence of a time-dependent surface charge is usually neglected entirely \cite{chem_time2, chem_time3, database_need_theory}. We are aware of one theoretical work \cite{koopal} and associated review \cite{koopalreview} that considers a surface charge that affects the rate constants of ion-association, which, however, does not consider the (chemistry-specific) non-linear dynamics induced by the electrostatic feedback as we do here.
		
		In this Letter we present a theory for the charging dynamics of solid surfaces. We include the Coulombic ion-surface interactions and reveal an intricate dependence on the reaction mechanism and the valency of the reactive ions already present in a mean-field description. The Coulomb interactions not only affect the time constant of the late-time exponential decay of the surface charge towards equilibrium after an ion concentration (or pH) shock, but they also induce strongly nonlinear dynamics at early times far from equilibrium.
		Combined with the present-day capability to experimentally measure the time-dependent surface charge density, our theory forms a first step to unveil the surface chemistry of technologically important but ill-understood materials \cite{database_need_theory, review_electrification_2}, such as silica \cite{silica1, 2pK_2} and graphene \cite{Graphene_BN}, and of processes such as the clean-up of radioactive and heavy metals \cite{soil1,database,heavymetalreview,cadmiumgoethite}. \\

    Surfaces, for instance silica, in water commonly charge either by desorption of ionic species from neutral surface groups or by adsorption of ionic species onto neutral surfaces. While the exact charging mechanism of the silica-water interface is complex, there is support for charging by desorption of protons at high pH and adsorption of protons at low pH  \cite{silica1, silica2, 2pK, 2pK_2},
    \begin{align}
		\ch{SiOH}_{\rm{(s)}}&\xrightleftharpoons[k_{\rm{a}}\rho]{k_{\rm{d \ }}}\ch{SiO}^-_{\rm{(s)}}+\ch{H}^+_{\rm{(aq)}}, \label{reacSi1}\tag{1a}	\\
		\ch{SiOH}_{\rm{(s)}}+\ch{H}^+_{\rm{(aq)}}&\xrightleftharpoons[k_{\rm{d \ }}]{k_{\rm{a}}\rho}\ch{SiOH_2}^+_{\rm{(s)}},\label{reacSi2}\tag{1b}
	\end{align}
	where $\ch{SiOH}_{\rm{(s)}}$ is a neutral silanol group that is covalently bound to the (solid) glass and where $\ch{SiO}^-_{\rm{(s)}}$ and $\ch{SiOH_2}^+_{\rm{(s)}}$ denote a silanol group with a proton desorbed or adsorbed in Eqs.~(\hyref{reacSi1}) and (\ref{reacSi2}), respectively. Here $\rho$ denotes the proton density at the solid surface, and the dissociation and association rate $k_{\rm{d}}$ and $k_{\rm{a}}$ will be discussed below for the charging kinetics of a single desorptive and a single adsorptive reaction, not only for monovalent reactive ions as in Eqs.~(\hyref{reacSi1}) and (\ref{reacSi2}) but for general valency $z$. While adsorption isotherms of real materials can rarely be described by just a single charging reaction \cite{2pK,database}, we show in Supplemental Material I (SM I \cite{SM}) that charging by multiple reactions can actually be well-approximated by the single-reaction kinetics presented in this Letter for a wide range of experimental conditions.
	
	We consider a macroscopic surface with a density $\Gamma$ of identical surface groups. A group can only be in either a neutral or a charged state. The charging is assumed to take place either by desorption (labeled by $-$) of a cation of charge $ze$, or by adsorption (labeled by $+$) of a cation of charge $ze$, with $z\geq 0$ and $e$ the proton charge. The surface densities of charged and neutral groups are denoted by $\sigma_\pm>0$ and $\Gamma-\sigma_\pm>0$ respectively, and the surface charge density is given by $\pm ze\sigma_\pm$.  Note that the charging dynamics is invariant under the sign of the reacting ions, and without loss of generality we can restrict attention to reactive cations of (strictly positive) valency $z$. Assuming the chargeable surface sites to be independent, we can describe the reaction kinetics in terms of the time-dependent surface density $\sigma_\pm(t)>0$ which satisfies Langmuir kinetics described by \cite{langmuir, bazant_account}
	\begin{equation}
	\partial_t\sigma_-=k_{\rm{d}}(\Gamma-\sigma_-)-k_{\rm{a}}\sigma_- \rho(\sigma_-)
	\label{Langmuira}
	\tag{2a}\\
    \end{equation}
    for desorptive charging reaction (\hyref{reacSi1}), and
    \begin{equation}
	\partial_t\sigma_+=k_{\rm{a}}(\Gamma-\sigma_+)\rho(\sigma_+)-k_{\rm{d}}\sigma_+
	\label{Langmuirb}
	\tag{2b}
	\end{equation}
    for adsorptive charging reaction (\hyref{reacSi2}). Here $k_\mathrm{d}$ and $k_\mathrm{a}$ are the rate constants of the dissociation and association of the reactive ion and $\rho(\sigma_\pm)$ is the volumetric concentration of reactive ions at the surface, which is defined at the position where the rate-limiting step for the reaction occurs \cite{BN2, bazant_account}. We consider this surface to be impermeable to non-reacting ions and therefore do not account for any Stern layer other than the charged surface groups \cite{chem_macr4}. The equilibrium surface charge follows from $\partial_t\sigma_\pm=0$ and is given by $\sigma_{\pm,\rm{eq}}=\Gamma(1+(k_{\rm{a}}\rho_{\rm{eq}}/k_{\rm{d}})^{\mp 1})^{-1}$, which reduces to an explicit ``Langmuir isotherm'' in the case that the equilibrium concentration of the reactive ions $\rho_{\rm{eq}}\equiv \rho(\sigma_{\pm,\rm{eq}})$ is a constant independent of $\sigma_{\pm,\rm{eq}}$. In general, however, this Langmuir isotherm is a self-consistency equation for $\sigma_{\pm,\rm{eq}}$ that requires an additional ``closure'' relation $\rho(\sigma_{\pm})$ for an explicit equilibrium solution $\sigma_{\pm,\rm{eq}}$. Without (Coulombic) interactions between surface and ions, the local concentration $\rho(\sigma_\pm)$ of reactive species in the vicinity of the surface would be equal to the bulk concentration $\rho_{\rm{b}}$ of the reactive ions far from the surface (which is independent of $\sigma_\pm$ and hence also independent from the reaction mechanism), such that Eqs.~(\hyref{Langmuira})-(\hyref{Langmuirb}) would be linear differential equations whose solution  can be written as $s_\pm(t)=1+(s_\pm(0)-1)\exp[-
	(k_{\rm{d}}+k_{\rm{a}}\rho_{\rm{b}})t]$ with the dimensionless charge $s_\pm=\sigma_\pm/\sigma_{\pm,\rm{eq}}$ such that $s_{\pm,\rm{eq}}=1$; here $s_\pm(0)-1$ is the integration constant and denotes the relative deviation from equilibrium at the initial time $t=0$. Note that the condition that $0\leq \sigma_\pm(t)\leq \Gamma$ implies that $0\textless s_\pm(0)\textless \Gamma/\sigma_{\pm,\rm{eq}}$, where the lower bound corresponds to an initially neutral surface whereas the upper bound can be as large as $\mathcal{O}(10-100)$, since typical equilibrium conditions have a charge occupancy of only a few percent of the total number of chargeable groups \cite{database_sites}.  Thus from measurements of $\sigma_\pm(t)$ at various concentrations of reactive (dissolved) species both $k_{\mathrm{d}}$ and $k_{\mathrm{a}}$ could in this non-interacting case be determined. 
	
	However, as the charged surface attracts or repels reactive ions, Eqs.~(\hyref{Langmuira}) and (\hyref{Langmuirb}) are complicated by a nontrivial relation $\rho(\sigma_\pm)$, which causes a charge-dependent decay rate and introduces deviations from purely single-exponential relaxation of $\sigma_\pm(t)$. In fact, an explicit function $\rho(\sigma_\pm)$ is needed to investigate and solve the dynamics, which we will develop here.
	\begin{figure}[t!]
			\includegraphics[width=.99\linewidth]{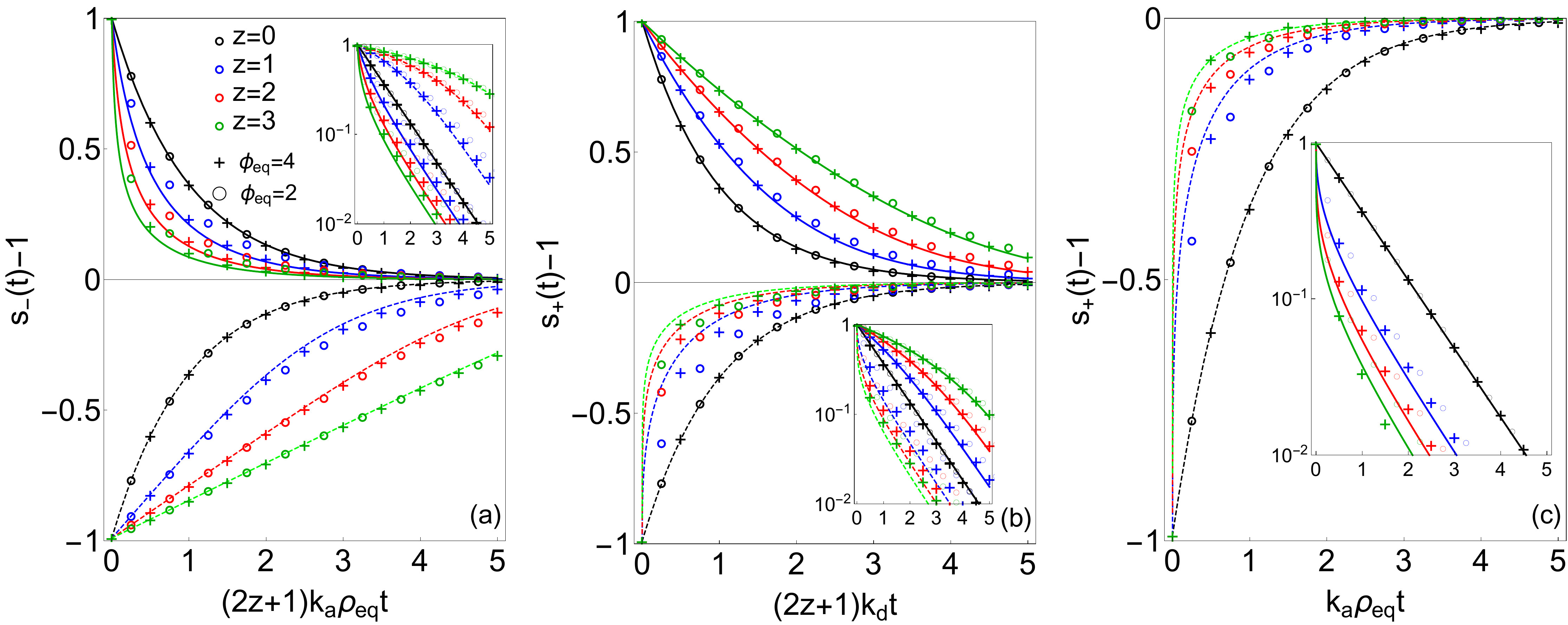}
		\caption{Time-dependent relative deviations $s_\pm(t)-1$ from the equilibrium charge density as follows from the kinetic Langmuir-Gouy-Chapman equations (\hyref{Langmuira})-(\hyref{3}) in (a) and Eqs.~(\hyref{Langmuirb})-(\hyref{3}) in (b) and (c), for equilibrium zeta potentials $(k_{\rm{B}}T/e)|\phi_{\rm{eq}}|$ equal to 50 mV (cross) and 100 mV (circle) for valencies $z=0,1,2,3$ (colors), in (a) $s_-(t)$ for desorptive reactions when $\sigma_{-,\rm{eq}}\ll\Gamma$, in (b) $s_+(t)$ for adsorptive reactions when $\sigma_{+,\rm{eq}}\ll\Gamma$, and in (c) $s_+(t)$ for adsorptive reactions when $\sigma_{+,\rm{eq}}\simeq \Gamma$. Insets denote semi-logarithmic representations of $|s_\pm(t)-1|$. The case $\sigma_{-,\rm{eq}}\simeq \Gamma$ (not shown) is trivial with single-exponential decay for all $z$. 
		}
		\label{fig:theoreticalcurves}
	\end{figure}
	We consider the planar and homogeneous chargeable solid surface discussed above in contact with a bulk solvent with permittivity $\epsilon$ and temperature $T$ with a three-component $1:1:z$ electrolyte of bulk concentrations $\rho_{\rm{s}}:(\rho_{\rm{s}}-z\rho_{\rm{b}}):\rho_{\rm{b}}$. For convenience we assume trace amounts of reactive ions and therefore set $\rho_{\rm{b}}\ll\rho_{\rm{s}}$, where $\rho_{\rm{s}}$ is the bulk salt concentration.  We also assume the electrolyte volume to be macroscopically large such that $\rho_{\rm{b}}$ and $\rho_{\rm{s}}$ do not change due to surface charging. Furthermore, we assume the charging timescale $\tau_\pm$,  which remains to be derived, to be the slowest timescale of the system. Given that the typical timescale for electric double layer (EDL) equilibration is around $10^{-9}-10^{-6}$ s and that the (geometry and flow dependent) transport timescale for ions in stirred reactors can be as short as $10^{-4}$ s \cite{transporttime}, we find a large window $\tau_\pm\gg 10^{-4}$ s for reactions to be well-described by our (reaction-limited) theory \cite{rate-limit}: for example phosphate desorption shows characteristic reaction timescales of hours \cite{phosphate} and adsorption of transition metals can occur on millisecond timescales \cite{pressure_jump, pressure_jump2}. The slow-reaction assumption allows us to describe the EDL within an equilibrium theory, for which we take the Gouy-Chapman solution of Poisson-Boltzmann (PB) theory for simplicity \cite{polymer, PBformulary}. Although PB theory is based on a mean-field assumption for a system of point ions, it is known that for all but the highest salt concentrations this theory is quite accurate for $1:1$ and even $1:2$ aqueous electrolytes \cite{PB1}, and we expect a similar accuracy for $1:1:z$ electrolytes in the limit $\rho_{\rm{b}}\ll\rho_{\rm{s}}$ of our interest. Within these assumptions the concentration of reactive ions at the surface is determined by a Boltzmann distribution $\rho(\sigma_\pm)=\rho_{\rm{b}}\exp[-z\phi(\sigma_\pm)]$, where $k_{\ch{B}}T\phi(\sigma_\pm)/e$ is the electric potential at the surface with a surface charge $\pm ze\sigma_\pm$, with $k_{\rm{B}}$ the Boltzmann constant. For desorptive charging the surface and ions have opposite charge and hence $z\phi(\sigma_-)<0$, while for adsorptive charging ions and surface have the same sign yielding $z\phi(\sigma_+)>0$. With this observation the Gouy-Chapman solution for a $1:1$ electrolyte, which is relevant here as $\rho_{\rm{s}}\gg\rho_{\rm{b}}$, gives $\phi(\sigma_\pm)=\pm 2\sinh^{-1}(z\sigma_\pm /\sigma^*)$\cite{polymer, PBformulary}, where $\sigma^*=(2\pi \lambda_\mathrm{B}\lambda_\mathrm{D})^{-1}$ with $\lambda_\mathrm{B}=e^2(4\pi\epsilon k_\mathrm{B}T)^{-1}$ the Bjerrum length of the solvent and $\lambda_\mathrm{D}=(8\pi\lambda_\mathrm{B}\rho_{\rm{s}})^{-\frac{1}{2}}$ the Debye screening length. Substituting the Gouy-Chapman potential in the Boltzmann factor yields
	\begin{equation}
	\rho(\sigma_\pm)=\rho_{\rm{b}}\exp[- z\phi(\sigma_\pm)]=\rho_{\rm{b}}\bigg(\frac{z\sigma_\pm}{\sigma^*}+\sqrt{1+\big(\frac{z\sigma_\pm}{\sigma^*}\big)^2}\bigg)^{\mp 2z},
	\label{3}
        \tag{3}
	\end{equation}
    where the exponent is positive for desorptive charging and negative for adsorptive charging. 
	Because Eq.~(\ref{3}) is reaction-mechanism dependent, explicit information on the charging mechanism can be deduced from the reaction kinetics as described by combining Eq.~(\ref{3}) with Eqs.~(\ref{Langmuira})-(\ref{Langmuirb}). 
 
	In order to investigate the influence of the Coulombic ion-surface interactions on the charging dynamics, we numerically solve $\sigma_-(t)$ from the kinetic Langmuir-Gouy-Chapman Eqs.~(\hyref{Langmuira})-(\hyref{3}). The symbols in Fig.\ref{fig:theoreticalcurves}(a) present the resulting relative deviations from equilibrium, $s_-(t)-1$, for a desorptive reaction in the experimentally common case of low equilibrium saturation $\sigma_{-,\rm{eq}}\ll\Gamma$, both for $s_-(t=0)=2$ and $0.01$ corresponding to a surface with double the charge compared to equilibrium and an initially almost uncharged surface, respectively, for equilibrium surface potentials of 50 mV ($|\phi_{\rm{eq}}|=2$, circles) and 100 mV ($|\phi_{\rm{eq}}|=4$, crosses) and for valencies $z=0,1,2,3$ indicated by the different colours.  Fig.\ref{fig:theoreticalcurves}(a) shows that a desorptive surface that is overcharged ($s_->1$) decays to equilibrium faster than one that is undercharged ($s_-<1$), the more so for larger valencies $z$. Interestingly, the sorption of uncharged species ($z=0$, black symbols) reveals perfect symmetry between the two cases as expected for first order kinetics, which is also manifest in the semi-logarithmic representation of $|s_-(t)-1|$ in the inset of Fig.\ref{fig:theoreticalcurves}(a) that shows a data collapse and a single-exponential decay for $z=0$. For $z\geq1$ the inset reveals a non-exponential time dependence with an initially slower decay for undercharged surfaces and an initially faster decay for overcharged surfaces, the difference becoming more pronounced for higher valencies.  Fig. \ref{fig:theoreticalcurves}(b) shows the deviation $s_+(t)-1$ from equilibrium for numerical solutions of Eqs.~(\hyref{Langmuirb})-(\hyref{3}) for an adsorptive charging reaction and the same low equilibrium surface density $\sigma_{+,\rm{eq}}\ll \Gamma$ and the same surface potentials and valencies as in (a). Interestingly, for this reaction the relaxation of an initially undercharged surface to equilibrium is faster, rather than slower as we found for desorptive undercharged surfaces in (a). Hence the two mechanisms can be distinguished by inspecting a single time-trace of the surface charge. We do not plot the dynamics of a desorptive surface that is saturated in equilibrium $\sigma_{-,\rm{eq}}\simeq \Gamma$ (in which case $k_{\rm{d}}\gg k_{\rm{a}}\rho_{\rm{eq}}$) as the equilibration (dissociation) rate for such a surface $\partial_ts\simeq -k_{\rm{d}}(s-1)$ is linear and equilibration occurs through trivial single-exponential decay. The lack of non-linearity for such a surface stems from the fact that the dissociation process is unaffected by the electrostatic surface-ion interaction. However, as can be seen in Fig.\hyref{fig:theoreticalcurves}(c) the dynamics of an adsorptively charged surface with a saturated charge density $\sigma_{+,\rm{eq}}\simeq \Gamma$ is markedly non-linear. As was the case in Fig.\hyref{fig:theoreticalcurves}(b) we see that an undercharged surface equilibrates faster than a single-exponential. Clearly, these rather distinctive features of the time-dependent surface charge contain explicit information on not only the reaction mechanism but also the valency of reacting ions. Interestingly, such deviations from single-exponential decay have historically been observed in pressure-jump experiments \cite{pressure_jump, pressure_jump2}. In SM II \cite{SM} we show that these experiments are well described by our theory, alleviating the need of introducing multiple reactions to describe such experiments.

	In Figs.\ref{fig:theoreticalcurves}(a)-(b) the dimensionless time on the horizontal axes contains a factor $(2z+1)$, which as we will show now, is convenient as it leads to a data collapse in the asymptotic nonlinear-screening regime $|\phi_{\rm{eq}}|\gg 1$ where $s_\pm(t)$ only depends on the valency, the reaction mechanism, and the initial charge state. To see why the near-equilibrium decay rate includes a factor $(2z+1)$ in Figs.\hyref{fig:theoreticalcurves}(a) and (b) but not (c) we simplify the Langmuir-Gouy-Chapman Eqs.~ (\hyref{Langmuira})-(\hyref{Langmuirb}) and (\hyref{3}) in the important and common case of large equilibrium surface potentials where $z\sigma/\sigma^*\textgreater 1$, say beyond 50 mV where $|\phi_{\rm{eq}}|\geq 2$. In this limit Eqs.~(\hyref{Langmuira})-(\hyref{Langmuirb}) can be rewritten as a single polynomial (Chini \cite{chini}) differential equation 
    \begin{subequations}
    \begin{gather}
    -\partial_t\label{dif eq delta s} s_\pm=k_{\rm{a}}\rho_{\rm{eq}}\bigg(s_\pm^{1\mp2z}-s_\pm^{-z\mp z}\bigg)+k_{\rm{d}}\bigg(s_\pm-s_\pm^{-z\mp z}\bigg),\tag{4}
    \end{gather}
    \end{subequations}
    for which a closed form solution can be obtained by separation of variables only for an adsorptively charged surface with $k_{\rm{d}}\gg k_{\rm{a}}\rho_{\rm{eq}}$ in which case $s^{2z+1}_+(t)=1+(s_+^{2z+1}(0)-1)\exp[-(2z+1)k_{\rm{d}}t]$. Near-equilibrium, $s_\pm\simeq 1$, Eq.~(\hyref{dif eq delta s}) simplifies to the linear differential equation $\partial_ts_\pm\simeq- (s_\pm-1)/\tau_\pm$ with the near-equilibrium decay rate for desorptive and adsorptive charging given, respectively, by
    \begin{align}
       \tau_-^{-1}&=(2z+1)k_{\rm{a}}\rho_{\rm{eq}} +k_{\rm{d}}\tag{5a},\\
       \tau_+^{-1}&=(2z+1)k_{\rm{d}} +k_{\rm{a}}\rho_{\rm{eq}}\tag{5b}.
    \end{align} 
    As announced, this timescale $\tau_\pm$ shows that electrostatic attraction can alter the linear, near-equilibrium, decay rate by a factor $(2z+1)$  for $z\geq 1$ compared to the neutral case ($z=0$) in the experimentally common regime $(k_{\rm{a}}\rho_{\rm{eq}}/k_{\rm{d}})^{\mp 1}\gg 1$ where $\sigma_{\pm,\rm{eq}}\ll \Gamma$. As for the majority of surfaces the equilibrium charge is much lower than saturation,  $\sigma_{\rm{eq}}\ll \Gamma$ \cite{database_sites}, we expect the correction by a factor $(2z+1)$ to be common and we note in passing that the only other work focusing on the influence of Coulombic ion-surface interactions on kinetics \cite{koopal, koopalreview} does not mention this factor. As already observed in Fig.\hyref{fig:theoreticalcurves}, our simple Eq.~(\hyref{dif eq delta s}) shows that far-from-equilibrium the dynamics becomes non-linear and importantly the $\pm$ sign of the reaction mechanism breaks the near-equilibrium symmetry of dynamics with regard to a charge excess or a charge deficit. As can be seen from Eq.~(\hyref{3}), desorptively charged surfaces which are overcharged, $s_->1$, will initially attract an excess of reactive ions to the surface, thereby having an increased reaction rate compared to uncharged equilibration. Hence the equilibration for large overcharging is faster than expected from uncharged Langmuir kinetics. Conversely, an initially undercharged surface, $s_-<1$, will have a shortage of reactive ions and thus equilibration will be slower. For adsorptively charged surfaces equilibration will be non-linear regardless of $\sigma_{\rm{eq}}/\Gamma$ and here undercharging leads to a shortage of reactive ions compared to equilibrium and hence faster equilibration, as can be seen in Fig.\hyref{fig:theoreticalcurves}(b). The rate changing during equilibration is reminiscent of autocatalytic reactions where the equilibration rate changes because a catalyst speeding up the reaction is produced simultaneously with a reaction product \cite{autocat, autocat2, autocat_jan}, and in SM III \cite{SM} we demonstrate the similarity between Eq.~(\hyref{dif eq delta s}) and autocatalytic kinetics. A characteristic feature of such autocatalytic reactions is an increasing decay rate up to a maximum and a corresponding inflection point in the time-dependent decay. 
    \begin{figure}[t!]
		\centering
		\includegraphics[width=0.9\linewidth]{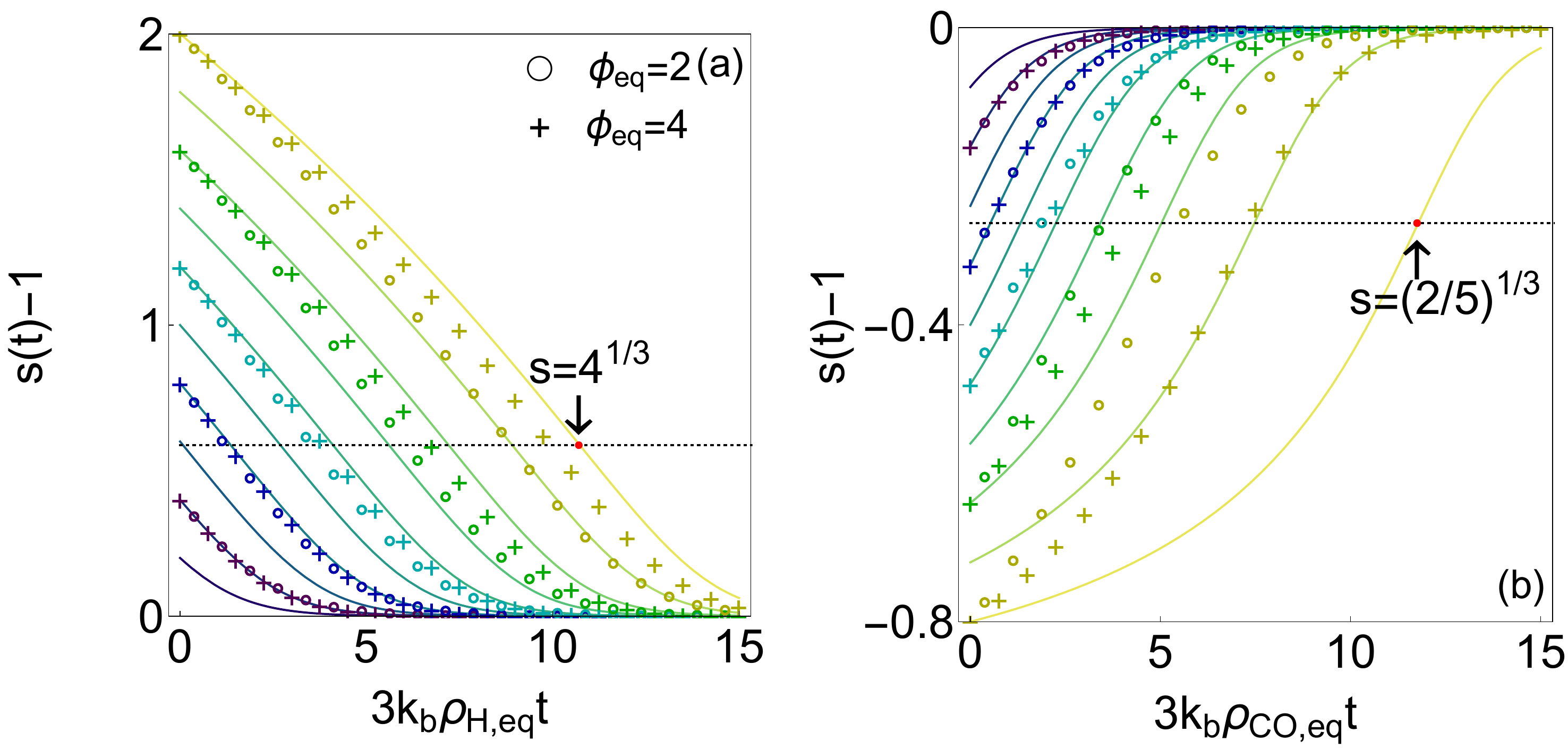}
		\caption{Time dependence of the relative deviation from the equilibrium charge density $s(t)-1$ for the autocatalytic ion displacement reaction Eq.~(\hyref{Ca_dif}) and Eq.~(\hyref{apatit_dif}) for 10 different initial conditions $s(0)$ in the experimentally common regime $\sigma_{\rm{eq}}\ll\Gamma$. Symbols are numerical solutions for the full Langmuir-Gouy-Chapman equation (SM IV \cite{SM}) for $\phi_{\rm{eq}}=2$  (open symbols) and $\phi_{\rm{eq}}=4$ (crosses), for 5 initial conditions. Gray dashed line denotes the inflection point (top $s(t)-1\simeq 0.6$, bottom $s(t)-1\simeq -0.25$).}
		\label{fig:S curves}
	\end{figure}
     Interestingly, for so-called ion displacement reactions in which ions are involved in both the forward- and the back-reaction, inflection points are easily realisable as there are now two ions attracted or repelled from the charged surface such that the Coulombic feedback is strengthened: inflection points are hence a smoking gun that multiple ions are involved in a charging reaction. An example of a two-ion reaction where all reacting ions are repelled from the charged surface is the calcium charging of silica \cite{silica1} of Eq.~(\hyref{Ca}), while an example of a reaction where all the reacting ions are attracted to the charged surface is the fluoride charging of the biomineral carbonato-apatite \cite{carb} of Eq.~(\hyref{apatit}),
\begin{align}
	\ch{SiOH}_{\rm{(s)}}+\ch{Ca}^{2+}_{\rm{(aq)}}&\xrightleftharpoons[k_{\rm{b }}\rho_{\rm{H}}]{k_{\rm{f}}\rho_{\rm{Ca}}} \ch{SiOCa}^+_{\rm{(s)}} +\ch{H}^+_{\rm{(aq)}}\label{Ca}\tag{6a},\\
		    \ch{X}\ch{CO_3}_{\rm{(s)}} + \ch{F}^{-}_{\rm{(aq)}}&\xrightleftharpoons[k_{\rm{b}}\rho_{\rm{CO}}]{k_{\rm{f}}\rho_{\rm{F}}} \ch{XF}^+_{\rm{(s)}} + \ch{CO_3}^{2-}_{\rm{(aq)}},\label{apatit}\tag{6b}
\end{align}
    where $\ch{X}=\ch{Ca_{10}(PO_4)_6}$, $k_{\rm{f}}$ is the forward (charging) reaction rate, $k_{\rm{b}}$ is the backward (decharging) rate and $\rho_i$ is the concentration of ion species $i$ at the charged surface. In SM III \cite{SM} we derive under the same  Gouy-Chapman and large surface potential conditions of the main text that the charge equilibration for Eq.~(\hyref{Ca}) and Eq.~(\hyref{apatit}) are respectively described by
	\begin{align}
	-\partial_t s & = k_{\rm{f}}\rho_{\ch{Ca,eq}}(s^{-3} -s^{-4})  +   k_{\rm{b}}\rho_{\rm{H,eq}}(s^{-1}-s^{-4})\label{Ca_dif}\tag{7a},\\
	 	-\partial_t s & = k_{\rm{f}}\rho_{\ch{F,eq}}(s^3-s^2) \ \ \ \ \ \  +  k_{\rm{b}}\rho_{\rm{CO,eq}}(s^5-s^2)\tag{7b},\label{apatit_dif}
	\end{align}
	with the resulting near-equilibrium decay constant $\tau=k_{\rm{f}}\rho_{i,\rm{eq}}+3k_{\rm{b}}\rho_{j,\rm{eq}}$ for both reactions. Comparing Eqs.~(\hyref{Ca_dif})-(\hyref{apatit_dif}) to Eq.~(\hyref{dif eq delta s}) we see that now the time evolution is given by the difference of two polynomials of (non-zero) unequal degree, ensuring that there is always a maximum in the decay rate and hence an inflection point. We find that reactions of the form (\hyref{apatit}) have two physically realizable inflection points located at $s=2/3$ (if $\sigma_{\rm{eq}}\simeq\Gamma$) and $s=(2/5)^{1/3}\simeq0.75$  (if $\sigma_{\rm{eq}}\ll\Gamma$): the reaction (\hyref{Ca}) has only one accessible inflection point $s=4^{1/3}\simeq1.6$ (if $\sigma_{\rm{eq}}\ll \Gamma$), while its second inflection point $s=4/3$ is inaccessible for a saturated surface with $\sigma_{\rm{eq}}\simeq \Gamma$.\\
 
	We plot the dynamics resulting from the Eqs.~(\hyref{Ca_dif})-(\hyref{apatit_dif}) for a variety of starting conditions $s(0)$ in Fig.\hyref{fig:S curves} in the experimentally common limit $\sigma_{\rm{eq}}\ll \Gamma$. In Fig.\hyref{fig:S curves}(a) excellent agreement between the asymptotic Eq.~(\hyref{Ca_dif}) and full numerical results can be seen, while in Fig.\hyref{fig:S curves}(b) for  large undercharging $s(0)\textless 0.2$ a discrepancy between Eq.~(\hyref{apatit_dif}) and the full numeric solution is found. However in general Eqs.~(\hyref{Ca_dif})-(\hyref{apatit_dif}) predict the location of the inflection point accurately for a range of common surface potentials $(k_{\rm{B}}T/e)|\phi_0|\in[50,100]$ mV. For ion-displacement reactions of the form (\hyref{Ca}) and (\hyref{apatit}) involving ions with higher valencies but monovalent surface charge the inflection point will lie even closer to equilibrium. Thus surfaces that are initially undercharged by only $\simeq25\%$ or overcharged by $\simeq60\%$ will generally equilibrate along sigmoidal curves, which is a distinguishing feature that cannot be observed for the single-ion reactions Eqs.~(\hyref{reacSi1})-(\hyref{reacSi2}). Finally we note that ion-displacement reactions offer a simple explanation for the recently observed sigmoidal equilibration of the surface charge at an aqueous silica interface \cite{autocat_jan} using only a single charging-reaction of the form Eq.~(\hyref{Ca}) rather than the proposed autocatalytic dissolution cycle involving multiple steps. 
    
	To summarize, in this Letter we present a model for the non-linear dynamics of reaction-limited surface charging, combining Langmuir dynamics with Poisson-Boltzmann theory. The model captures how the screened electrostatic surface-ion interaction affects the reaction rate near and far from equilibrium in terms of a non-linear differential equation, where the electrostatic interaction is described by only using the charge valency of the reactive ion. The Coulombic ion-surface interaction leads to a charge-dependent decay rate, which can be used to gain information on the valency of reacting ions,  initial charge, and reaction rate. De- and adsorptive reactions can be distinguished by inspecting whether far-from-equilibrium decay is slower or faster than near-equilibrium decay, while an inflection point is characteristic for two-ion reactions. Interestingly, we note that inflection points are a characteristic feature of autocatalytic reactions and that the electrostatic ion-surface interaction can be seen as a catalytic feedback loop. Hence electrostatics offers a straightforward explanation for the recently measured autocatalytic charging of silica \cite{autocat_jan}.
    \begin{acknowledgments}
	We thank Ben Werkhoven and Cheng Lian for fruitful discussion, and an anonymous referee for pointing us to the pressure-jump experiments Ref.~\cite{pressure_jump, pressure_jump2}. This work is part of the D-ITP consortium, a program of the Netherlands Organisation for Scientific Research (NWO) that is funded by  the  Dutch  Ministry  of  Education,  Culture  and  Science (OCW).
    \end{acknowledgments}

\end{document}


\title{Supplemental Material for ``Coulombic surface-ion interactions induce non-linear and chemistry-specific charging kinetics''}
\author{W.Q. Boon}
	\affiliation{Institute for Theoretical Physics, Utrecht University,  Princetonplein 5, 3584 CC Utrecht, The Netherlands}
	\author{M. Dijkstra}
	\affiliation{Soft Condensed Matter, Debye Institute for Nanomaterials Science, Utrecht University, Princetonplein 1, 3584 CC Utrecht, The Netherlands}
	\author{R. van Roij}
		\affiliation{Institute for Theoretical Physics, Utrecht University,  Princetonplein 5, 3584 CC Utrecht, The Netherlands}
	    \maketitle
    \section{When can multiple reactions be described by single-reaction kinetics?}

    The single-reaction (one-pKa) charging reaction presented in this Letter is a simplified representation for many real liquid-solid interfaces, which often require multiple surface charging reactions for the reproduction of measured equilibrium Langmuir isotherms\cite{2pK, 2pK_2,database, review_electrification_2}. In this Supplemental Material we show that for a wide range of conditions a two-reaction model is well approximated by the single-reaction model presented in this Letter. For simplicity we will consider a two-reaction (two-pKa) system with two distinct surface-sites labeled $\ch{(SiOH)}_1$ and $\ch{(SiOH)}_2$, charged by similar desorptive surface reactions but with different ad- and desorption rates  
    \begin{subequations}
        \begin{gather}
        \ch{(SiOH)}_1 \xrightleftharpoons[k_{\rm{a1}}\rho]{k_{\rm{d1 \ }}} \ch{(SiO^-)}_1 +\ch{H^+}_{\rm{(aq)}},\tag{S1a}\\
        \ch{(SiOH)}_2 \xrightleftharpoons[k_{\rm{a2}}\rho]{k_{\rm{d2 \ }}}  \ch{(SiO^-)}_2 +\ch{H^+}_{\rm{(aq)}}\tag{S1b}, 
    \end{gather}
    \end{subequations}
  and hence the sites have different equilibrium areal densities $\sigma_{\rm{eq},1}\neq\sigma_{\rm{eq},2}$ if the site densities $\Gamma_i$ or equilibrium constants $K_i=k_{\rm{a},i}\rho/k_{\rm{d},i}$ are unequal. The Langmuir equation describing their charging kinetics is
        \begin{subequations}
        \begin{gather}
        \partial_t\sigma_1=k_{\rm{d}1}(\Gamma_1-\sigma_1)-k_{\rm{a}1}\rho(\sigma_e)\sigma_1,\tag{S2a}\label{S2a}\\
        \partial_t\sigma_2=k_{\rm{d}2}(\Gamma_2-\sigma_2)-k_{\rm{a}2}\rho(\sigma_{\rm{e}})\sigma_2,\tag{S2b}\label{S2b}
            \end{gather}
         \end{subequations} with the resulting areal charge density $\sigma_{\rm{e}}=\sigma_1+\sigma_2$ coupling the two reactions. As both reactions have the same reactive ion (H$^+$) a concentration change $\Delta\rho$ at the surface would cause the equilibrium to shift for both reactions, and the equilibration kinetics resulting after this concentration shift would in principle need to be described by the two coupled non-linear differential equations (\hyref{S2a}) and (\hyref{S2b}) as they are coupled by the Gouy-Chapman ``closure'', given by Eq.~(3)  in the Letter.         \begin{figure}[t!]
		\centering		
		\includegraphics[width=0.9\linewidth]{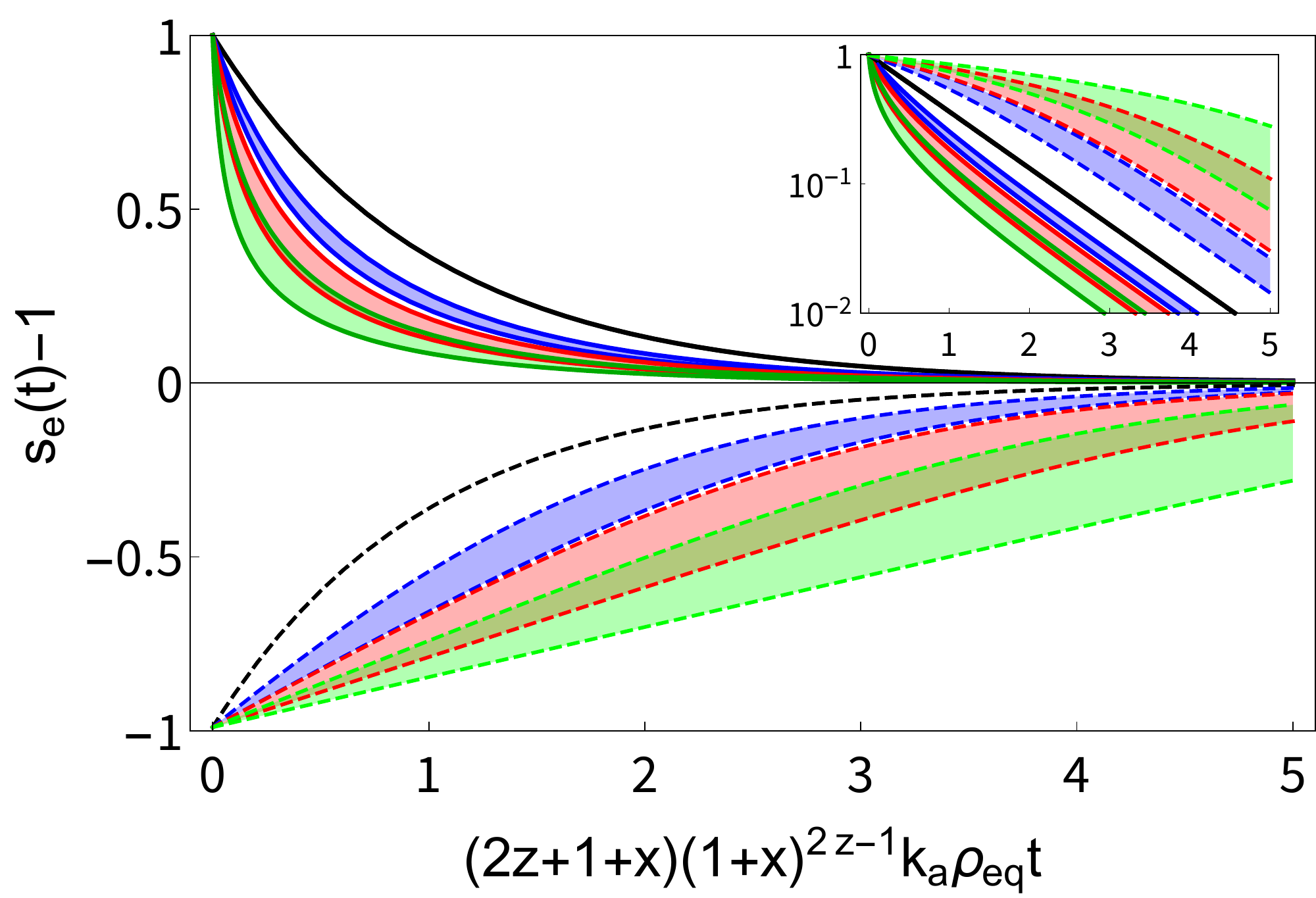}
		\caption{Replotting of Fig.1(a) showing charge relaxation for ion valencies $z=0,1,2,3$ (respectively black, blue, red, green) calculated with curves not originating from Eq.~(4) in the main Letter but instead plotted using Eq.~(\hyref{background charge}) with dimensionless static background-charge $x\in[0,1]$. The outermost curves with $x=0$ revert exactly to those described by Eq.~(4) in the main Letter, while curves with $x=1$ lie significantly closer to the black curve showing single-exponential relaxation, with intermediate $x$ lying in the shaded region. Note that the near-equilibrium decay rate has changed from $(2z+1)k_{\rm{a}}\rho_{\rm{eq}}$ to $(z+1+x)(1+x)^{2z-1}k_{\rm{a}}\rho_{\rm{eq}}$ which for $z=3$ and $x=1$ yields a factor 32 difference in the near-equilibrium decay rate. In general however for small $x$ the charging kinetics is well described by the single reaction of Eq.~(4) in the main Letter.}
		\label{fig:background}
	\end{figure}
    However, here we show that if the two reactions have dissimilar equilibrium constants $K_i$ or site densities $\Gamma_i$ one of the two differential equations can almost always be neglected, as when the equilibrium constants differ so does the shift in equilibrium charge density $\Delta\sigma_i=\sigma_{\rm{eq},i}(\rho+\Delta\rho)-\sigma_{\rm{eq},i}(\rho)$ for a given concentration shock $\Delta\rho$. When the difference in equilibrium charge densities of the two sites is very unequal (either $\Delta\sigma_1\gg\Delta\sigma_2$ or $\Delta\sigma_1\ll\Delta\sigma_2$) the change in surface charge can be described using a single reaction model (as $\partial_t\sigma_{\rm{e}}\simeq \partial_t\sigma_{\rm{1}}$ or $\partial_t\sigma_{\rm{e}}\simeq \partial_t\sigma_{\rm{2}}$). Solving for the coupled-Langmuir kinetics Eq.~(\hyref{coupled langmuir}) in steady-state, $\partial_t\sigma_i=0$, we find for $\sigma_{i,\rm{eq}}\ll\Gamma_i$ that the relative change in the equilibrium surface charge $\Delta\sigma_1/\Delta\sigma_2\propto (\Gamma_1 K_1)/(\Gamma_2K_2)$ while for nearly saturated surfaces with $\sigma_{\rm{eq}}\approx\Gamma$ the relative shift scales as $\Delta\sigma_1/\Delta\sigma_2\propto (\Gamma_1 K_2)/(\Gamma_2 K_1)$. As long as $\Delta\sigma_1/\Delta\sigma_2$ deviates significantly from unity, the shift in one of the two equilibrium-densities can be neglected for the shift in the total charge density. To find an explicit maximum bound for which this approximation is valid we calculate the maximum concentration shock $\Delta\rho_{\rm{m}}\in \rho[-1,\infty]$ for which this approximation holds by solving for $\Delta\sigma_1/\Delta\sigma_2=1$. This has a trivial solution $\Delta\rho=0$ when $\Gamma_1=\Gamma_2$ and $K_1=K_2$ and a non-trivial solution 
    \begin{equation}
    \begin{split}
        \frac{\Delta\rho_{\rm{m}}}{\rho}= \frac{1}{K_1K_2\rho^2} \frac{\Gamma_1 K_1\rho-\Gamma_2 K_2\rho}{\Gamma_2 (1+K_1\rho)-\Gamma_1(1+K_2\rho)}
        -\\
        \frac{\Gamma_1(2+K_2\rho)-\Gamma_2(2+K_1\rho)}{\Gamma_1(1+K_2\rho)-\Gamma_2(1+K_2\rho)},
        \end{split}\tag{S3}
        \label{range}
    \end{equation}
    which simplifies to $\Delta\rho_{\rm{m}}/\rho=(K_1K_2\rho^2)^{-1}-1$ in the case that $\Gamma_1\simeq \Gamma_2$. In general, the range of validity of Eq.~(\hyref{range}) is large when $K_1$ and $K_2$ are very unequal, except close to a concentration where $\rho^2K_1K_2=1$ where $\Delta\rho_{\rm{m}}$ tends to zero. Interestingly, in this case the charge density is exactly half-occupied $\sigma_{e}= (\Gamma_1+\Gamma_2)/2$, which is rare for most experimental conditions. For all other concentrations $\rho$ a very large range $\Delta\rho_{\rm{m}}/\rho$ of concentration shocks remains over which a two-reaction system essentially equilibrates through a single charging reaction. However, while the dynamics will be governed by a single reaction, there will be a static background charge due to which $\sigma_1\neq\sigma_{\rm{e}}$ such that the single reaction is still not exactly equal to the single-reaction kinetics in the Letter. That the deviation from the single-reaction kinetics in the Letter due to this static background charge is minor will be shown in the next paragraph.\\

    As the total surface charge density (in units of the elementary charge) is given by $\sigma_{\rm{e}}=\sigma_1+\sigma_2$, the Gouy-Chapman relation (Eq.~(3) in the Letter) between the surface occupancy $\sigma_1$ and concentration $\rho$ for two reactions now reads
    \begin{equation}
        \rho(\sigma_1)=\bigg(\frac{z(\sigma_1+\sigma_2)}{\sigma^*}+\sqrt{1+\bigg(\frac{z(\sigma_1+\sigma_2)}{\sigma^*}\bigg)^2}\bigg)^{2z}.\tag{S4}
        \label{conc_alt}
    \end{equation}
    As discussed in the previous section we will now assume that $\Delta\rho\ll\Delta\rho_{\rm{m}}$ and without loss of generality we identify $\sigma_{\rm{eq},2}$  as the static-charge density ($\Delta\sigma_1/\Delta\sigma_2\gg1$) from which follows $\sigma_2\simeq\sigma_{\rm{eq},2}(\rho)\simeq\sigma_{\rm{eq},2}(\rho+\Delta\rho)=\mathrm{cnst}$. The non-dimensional change in total surface charge will then be $\partial_ts_{\rm{e}}=\partial_t\sigma_{\rm{e}}/\sigma_{\rm{eq,e}}\simeq\partial_t\sigma_1/\sigma_{\rm{eq,e}}$ and when $\sigma_{\rm{eq,e}}\gg\sigma^*$ by combining Eq.~(\hyref{S2a}) and Eq.~(\hyref{conc_alt}) we find 
    \begin{equation}
    \tag{S5}
    \begin{split}
    -\partial_ts_{\rm{e}}=k_{\rm{d}1}(s_{\rm{e}}-1)+k_{\rm{a}1}\rho_{\rm{eq}} \big(s_{\rm{e}}(s_{\rm{e}}+x)^{2z}-(1+x)^{2z}\big),
    \label{background charge}
    \end{split}
    \end{equation}
    where the dimensionless static surface charge density $x=\sigma_{\rm{eq,2}}/\sigma_{\rm{eq},e}\in [0,1]$ is the ratio of static charge $\sigma_{\rm{eq,2}}(\rho+\Delta\rho)$ \emph{after} the concentration shock. Interestingly, the near-equilibrium decay rate $\tau_-$ is altered from $(2z+1)k_{\rm{a}}\rho_{\rm{eq}}$ to $(2z+1+x)(1+x)^{2z-1}k_{\rm{a}}\rho_{\rm{eq}}$ and hence the deviation from uncharged Langmuir kinetics becomes even larger if the background surface charge is included. Clearly  both the timescale and dynamics revert back to the single-reaction kinetics of Eq.~(4) in the Letter when $x\ll 1$. To check how much the dynamics is affected at intermediate $x$ we replot Fig.1(a) in the Letter but now with Eq.~(\hyref{background charge}) with $z=0,1,2,3$ (black, blue, red, green) and $s=1$ and $s=-0.99$ instead of Eq.~(4). The resulting shaded regions in Fig.\hyref{fig:background} represent curves with different $x\in[0,1]$. We observe that for increasing $x$ the dynamics moves closer to single-exponential decay (black line), with this shift being more pronounced for increasing $z$. However, even for $x=1$ there is no over-dramatic difference from the single-reaction kinetics explored in the Letter and in many cases a two-reaction system is well approximated by the single-reaction model in the Letter. However one should be cautious when extracting the ion valency from dynamics around $x\simeq 1$ as here the dynamics closely resembles that of a single ion with $z-1$. We expect this may occur in processes such as the adsorption of heavy metal ions from ground water which occurs on pre-charged substrates \cite{database}. Under these conditions surface charging can be readily described using Eq.~(\hyref{background charge}).\\
    
    To summarize, here we have shown that the one-pKa charging reaction is a valid approximation for more complex systems, involving multiple charging reactions, when (i) the ratio of the equilibrium constant $K_1/K_2$ is not close to unity and (ii) the concentration shock $\Delta\rho$ is constrained within a range $\Delta\rho_{\rm{m}}$ which we show to be generally large.

    \section{Surface charging kinetics from pressure-jump experiments}
    While experimental investigations of surface charging kinetics are rare, several kinetic studies employing a pressure-jump technique exist \cite{pressure_jump, pressure_jump2}. In such an experiment a mixture of colloidal particles and aqueous electrolytes is slowly pressurized to more than $1$ MPa, thereby shifting the surface reactions at the colloidal surface to a high-pressure equilibrium. When this pressure is suddenly released, the solution pressure converges to atmospheric pressure in $\simeq$ 0.1 ms. After this jump, the colloidal surface charge must revert from its high-pressure equilibrium to its atmospheric equilibrium. The change in surface charge is measured indirectly, by using the solution conductance as a proxy for the surface charge. While it is unclear what the exact relation between charge and conductance is, a linear relation is often assumed \cite{pressure_jump2}.\\

\begin{figure}[t!]
		\centering		
		\includegraphics[width=0.95\linewidth]{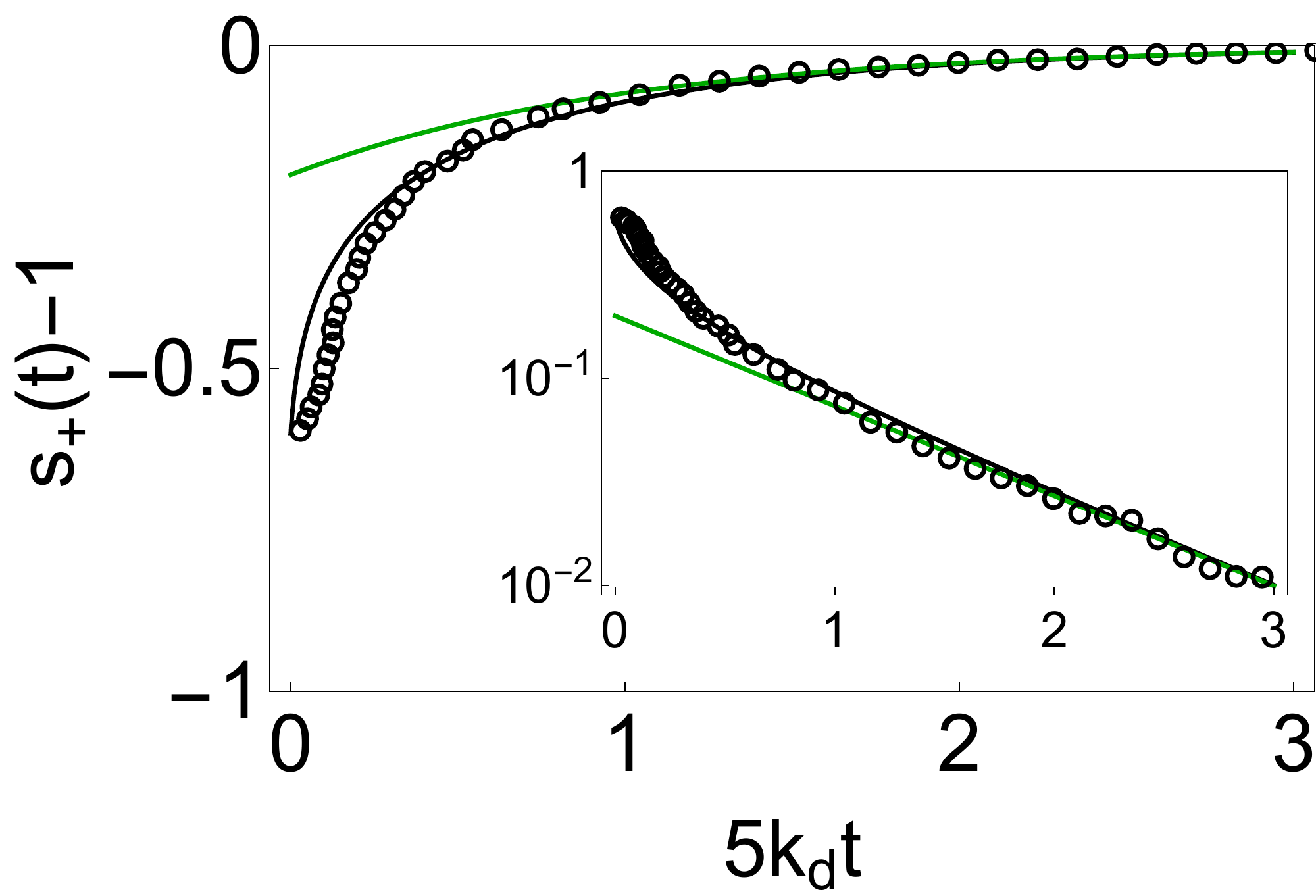}
		\caption{Comparison between experimental data from Fig.1 of Ref.~\cite{pressure_jump} (symbols), single exponential decay (green) and non-linear dynamics (black, Eq.~(\hyref{eq jump}) with initial degree of undercharging $s(0)-1=-0.6$. The reaction time $\tau_+\simeq 27$ ms is fitted from the late time decay, yielding $k_{\rm{d}}^{-1}\simeq 135$ ms. Inset shows the same data in a semi-logarithmic representation. }
		\label{fig:jump}
	\end{figure}

    Of particular interest is a set of experiments where the adsorption of divalent transition metals such as Cu$^{2+}$, Pb$^{2+}$, Mn$^{2+}$, and Co$^{2+}$  onto $\gamma$-alumina (Al$_2$O$_3$) particles is studied. Here the authors reject simplest adsorption mechanism, for Cu$^{2+}$ given by 
    \begin{equation}
    \rm{AlOH}_{\rm{(s)}}+\rm{Cu}^{2+}_{\rm{(aq)}}\xrightleftharpoons[k_{\rm{d \ }}]{k_{\rm{a}}\rho} AlOHCu^{2+}_{\rm(s)},\tag{S5}
    \label{reac jump}
    \end{equation}
    because this single-step reaction does not show single-exponential decay \cite{pressure_jump}. Their theoretical model used for analysis fixes the surface (zeta) potential, and hence they find that multiple reactions are needed to explain the observed charge equilibration \cite{pressure_jump, pressure_jump2, relaxation_kinetics}. While their analysis is valid for fixed surface potential, for a potential varying with surface charge we actually expect deviations from single-exponential decay, as shown by Eq.~(4) in our Letter. The Langmuir-Gouy-Chapman dynamics of reaction (\hyref{reac jump}) with $z=2$ for surface potentials larger than 50 mV is given by
    \begin{equation}
    -\partial_t s_+=k_{\rm{a}}\rho_{\rm{eq}}\bigg(s_+^{-3}-s_+^{-4}\bigg)+k_{\rm{d}}\bigg(s_+-s_+^{-4}\bigg).\tag{S6}
    \label{eq jump}
    \end{equation}
    To test whether the non-linear dynamics observed in the pressure-jump experiment are explained by Eq.~(\hyref{eq jump}), we extract the experimental data from the relevant pressure-jump experiment for copper adsorption (Fig.1 in Ref.~\cite{pressure_jump}). In this experiment it is found that the conductivity after the pressure jump decreases and reaches a constant value within 200 ms. In Fig.\hyref{fig:jump} we compare the experimental data (symbols)  with equilibration expected from Eq.~(\hyref{eq jump}) (black line), where we assume the experimentally common case where $\sigma_{\rm{eq}}\ll \Gamma$ and extract the reaction time $\tau_+= (5k_{\rm{d}})^{-1}\simeq 27$ ms from the single-exponential, late time, relaxation (green line). As the final equilibrium charge density is not measured in a pressure-jump experiment we use the degree of undercharging $s(0)-1\simeq -0.6$ as a fit parameter, where we note that an undercharged surface naturally explains why conductivity decreases: during equilibration mobile copper ions are taken out of solution. It can be seen that the difference between the single exponential decay (green) and experimental data (symbols) is large, but that Eq.~(\hyref{eq jump}) (black) can naturally explain a large part of the deviation from single-exponential decay, without needing to introduce a second reaction with a different timescale. \\
    
    While further analysis is required to reinstate reaction mechanism (\hyref{reac jump}), our analysis shows the importance of taking Coulombic surface-ion interactions into account when considering kinetics. Not only do we show that non-linear decay is readily captured by our model, also the desorption time $k_{\rm{d}}^{-1}$ is five times slower than the late-time decay $\tau_+$, in this divalent case. Furthermore, we have demonstrated that chemical information can easily be extracted from a single equilibration curve, even when the initial degree of undercharging is unknown. 
    
    \section{Similarity between surface charging and autocatalytic kinetics}
    Here we we will show the similarity between classical autocatalytic kinetics and surface charging as described by Eq.~(4) in the Letter. First we derive an approximate solution for this Chini differential Eq.~(4) in the main text, by expanding it up to second order around $s=1$ obtaining a Bernoulli differential equation (Supplementary Ref.\cite{Bernoulli}). Solving this equation by standard methods we obtain
    \begin{equation}
    \begin{split}
        & s_\pm(t)-1=\\
        & \dfrac{(s_\pm(0)-1)e^{-t/\tau_\pm}}{1\mp z \big(1- e^{-t/\tau_\pm}\big)\bigg(s_\pm(0)-1\pm \dfrac{s_\pm(0)-1}{(2z+1) (k_{\rm{d}}/k_{\rm{a}}\rho_{\rm{eq}})^{\pm 1}+1}\bigg)},
    \end{split}
        \label{Bernoulli}\tag{S7}
    \end{equation}
    where $\tau_\pm$ is given by Eq.~(5) in the Letter. Our Eq.~(\hyref{Bernoulli}) reverts to single exponential decay when $(s_\pm(0)-1)z\ll 1$, and is valid only as long as $\mp z(s_\pm(0)-1)\gg -1$. For most practical purposes Eq.~(\hyref{Bernoulli}) is not of much use, however it is interesting to note that a similar solution exists for autocatalytic equations thereby substantiating the claim that surface charging is autocatalytic. To make this comparison explicit we consider the simplest possible autocatalytic reaction (Eq.~(1a') from Ref.\cite{autocat2}),
    \begin{equation}
        \ch{A}_{(\rm{aq})}+\ch{Y}_{(\rm{aq})} \xrightleftharpoons[k_{\rm{b}}\rho_{\rm{Y}}^q]{ k_{\rm{f}}\rho_{\rm{A}}\rho_{\rm{Y}}} q \ \ch{Y}_{(\rm{aq})}\tag{S8}
    \end{equation}
    where the aqueous reactant $\ch{Y}_{(\rm{aq})}$ together with reactant $\ch{A}_{(\rm{aq})}$ produces $q$ copies of itself. The reaction is autocatalytic when the autocatalytic order $q\geq 2$. When the concentration $\rho_{\rm{A}}$ is constant the production rate of $\ch{Y}_{(\rm{aq})}$ is given by the Chini differential equation
    \begin{equation}
        -\partial_ty= k_{\rm{f}}\rho_{\rm{A}}(y^q-y),\tag{S9}
        \label{autocat_simple}
    \end{equation}
    with $y=\rho_{\rm{Y}}/\rho_{\rm{Y,eq}}$, which already shows similarities to Eq.~(4) in the Letter. To obtain a solvable Bernoulli equation we expand Eq.~(\hyref{autocat_simple}) up to second order around $x=1$, which has the solution
    \begin{equation}
        y(t)-1= \dfrac{(y(0)-1) e^{-t/\tau_{\rm{y}}}}{1-\dfrac{q}{2}(y(0)-1)(1-e^{-t/\tau_{\rm{y}}})},\label{S13}\tag{S10}
    \end{equation}
    with $\tau_{\rm{y}}=(q-1)k_{\rm{f}}\rho_{\rm{A}}$. While Eq.~(\hyref{S13}) is already very similar to Eq.~(\hyref{Bernoulli}), the similarity becomes even more apparent when comparing the autocatalytic dynamics to the dynamics of an adsorptively charged surface with $\sigma_{+,\rm{eq}}\ll\Gamma$ (where $k_{\rm{d}}\gg k_{\rm{a}}\rho_{\rm{eq}}$) in which case Eq.~(\hyref{Bernoulli}) simplifies to
    \begin{equation}
        s_+(t)-1= \dfrac{(s_+(0)-1) e^{-t/\tau_{+}}}{1-z(s_+(0)-1)(1-e^{-t/\tau_{+}})}.\tag{S11}\label{S14}
    \end{equation}
    Comparing Eq.~(\hyref{S13}) with Eq.~(\hyref{S14}) we find the only difference is the definition of the timescale $\tau_i$ and that the ion-valency $z$ replaces the autocatalytic order $q/2$. This correspondence between the ion valency $z$ and autocatalytic order $q$ supports the interpretation that the Coulombic ion-surface interactions acts autocatalytically.
    
     \section{Charging dynamics of ion displacement reactions}
    Here we will generalize the derivation of Eq.~(4) in the main text from single-ion reactions to two-ion reaction also known as ion-displacement reactions, yielding Eq.~(7) in the main text. We will consider a general form for the single-step ion-displacement reaction where an aqueous ion $\rm{A}_{\rm{(aq)}}^{\zbar_{\rm{A}}}$ displaces from the (uncharged) surface group $\rm{SB}_{\rm{(s)}}$ the ion $\rm{B}_{\rm{(aq)}}^{\zbar_{\rm{B}}}$, leaving a charged surface site $\ch{SA}^{\zbar_\ch{A}-\zbar_\ch{B}}_{\rm{(s)}}$ combining into the reaction
    \begin{equation}\ch{SB}_{\rm{(s)}}+\ch{A}^{\zbar_\ch{A}}_{\rm{(aq)}}\xrightleftharpoons[k_{\rm{b}}\rho_{\rm{B}}]{k_{\rm{f}}\rho_{\rm{A}}} \ch{S}^{\zbar_\ch{A}-\zbar_\ch{B}}_{\rm{(s)}}+\ch{B}^{\zbar_\ch{B}}_{\rm{(aq)}},\tag{S12}
    \end{equation}
    where $\zbar_\ch{A}$ and $\zbar_{\rm{B}}$ are the valencies of ad- and desorbing ions A and B respectively , which contrary to the ion-valency $z_i$ in the main text is here not considered to be strictly positive as the total charge difference between A and B is important. The resulting charge of the surface groups is given by $\zbar_\ch{S}=\zbar_\ch{A}-\zbar_\ch{B}$, such that the surface charge is $e\sigma \zbar_\ch{S}$. As in the main text, the charging dynamics will be described by Langmuir kinetics, which assumes identical and independent surface sites such that
    \begin{equation}
        \partial_t\sigma=k_\ch{A}\rho_\ch{A}(\sigma)(\Gamma-\sigma)-k_\ch{B}\rho_\ch{B}(\sigma)\sigma.\tag{S13}
    \end{equation}
  Note that now both terms contain the non-trivial $\rho(\sigma)$ dependence, which allows for sigmoidal equilibration as discussed in the Letter. For $\rho(\sigma)$ we use the Gouy-Chapman expression
    \begin{equation}
\rho_i(\sigma)=\rho_{\rm{b},\mathit{i}}\bigg(\frac{z_{\rm{S}}\sigma}{\sigma^*}+\sqrt{1+(\frac{z_{\rm{S}}\sigma}{\sigma^*})^2}\bigg)^{\pm 2z_i},\tag{S14}
    \end{equation}
  where $z_i=|\zbar_i|$. The exponent for $\rho_{\rm{i}}$ is positive when $\zbar_i\zbar_\ch{S}<0$ and negative when $\zbar_i\zbar_\ch{S}>0$. When $\zbar_i\zbar_\ch{S}=0$ the dynamics revert to the single-ion charging reaction in the main text. In the high charge limit $|\phi_{\rm{eq}}|>2$ (when  $z_{\rm{S}}\sigma_{\rm{eq}}/\sigma^*>1$) we find 
    \begin{equation}
    \begin{aligned}
        -\partial_ts=k_\ch{A}\rho_{\ch{A,eq}}  (s^{1\pm2z_\ch{A}}-s^{\pm2z_\ch{A}})+k_\ch{B}\rho_{\ch{B},\ch{eq}}\big(s^{1\pm2 z_{\ch{B}}}-s^{\pm2z_\ch{A}}\big),
        \end{aligned}\tag{S15}
    \end{equation}
    where the $\pm$ sign in front of $z_{\rm{i}}$ is negative when $\zbar_i\zbar_\ch{S}>0$ and positive when $\zbar_i\zbar_\ch{S}<0$. Substituting $\zbar_{\ch{A}}=+2$ and $\zbar_\ch{B}=+1$ (hence $\zbar_{\ch{S}}=+1$) we find Eq.~(7a) in the main text, while substituting $\zbar_\ch{A}=-1$ and $\zbar_\ch{B}=-2$ (hence $\zbar_{\ch{S}}=+1$) yields Eq.~(7b). Furthermore, for $\zbar_{\rm{A}}=0$ this equation reverts to desorptive charging and for $\zbar_{\rm{B}}=0$ this equation reverts to adsorptive charging as described in the Letter. The presented derivation can be naturally extended to single-step reactions involving an arbitrary number of charged species.